\documentclass[%
reprint, 
showpacs,
amsmath,amssymb,
aps,
prb,
floatfix,
lengthcheck,
]{revtex4-1}

\usepackage[pdftex]{graphicx} 
\usepackage{dcolumn}  
\usepackage{bm}       
\usepackage[usenames,dvipsnames]{color}
\definecolor{URLCOL}{rgb}{0,0.52,0.83} 
\definecolor{LINKCOL}{rgb}{0.05,0.5,0} 
\definecolor{orange}{rgb}{0.6,0.3,0} 
\definecolor{CITECOL}{rgb}{0.25,0,0.48} 
\usepackage{epstopdf}

\usepackage[caption=false]{subfig} 

\DeclareMathAlphabet{\gcal}{OMS}{cmsy}{m}{n}

\newcommand{\Lfun}[2]{{\gcal L}^{(#1)}_{#2}}

\begin{document}

\title{Emergence of Fourier's law of heat transport in quantum electron systems}

\author{Sosuke Inui}
\affiliation{Department of Physics, University of Arizona, 1118 East Fourth Street, Tucson, AZ 85721}
\affiliation{Department of Physics, Osaka City University, Sugimoto 3-3-138, Sumiyoshi-Ku, Osaka, 558-8585, Japan}

\author{Charles\ A.\ Stafford}
\affiliation{Department of Physics, University of Arizona, 1118 East Fourth Street, Tucson, AZ 85721}

\author{Justin\ P.\ Bergfield}
\affiliation{Department of Physics, Illinois State University, Normal, IL, 61761, USA}
\affiliation{Department of Chemistry, llinois State University, Normal, IL, 61761, USA}

\date{\today}
\begin{abstract}
The microscopic origins of 
Fourier's venerable law of thermal transport in quantum electron systems has remained somewhat of a mystery, 
given that previous derivations were forced to invoke intrinsic scattering rates 
far exceeding those occurring in real systems.  We propose an alternative hypothesis, namely, that 
Fourier's law emerges naturally if many quantum states participate in the transport of heat across the system.  
%
%
We test this hypothesis systematically in a graphene flake junction, and show that the temperature distribution becomes nearly classical when the 
broadening of the individual quantum states of the flake exceeds their energetic separation.
We develop a thermal resistor network model to investigate the scaling of the sample and contact thermal resistances, and show that the latter
is consistent with classical thermal transport theory in the limit of large level broadening. 
\end{abstract}

\maketitle

\section{Introduction}
\vspace{-.4cm}
As the dimensions of an electronic device are reduced, the power consumption, and concomitant heat generation, increases.  
Therefore, a detailed understanding of heat transport at the nanoscale 
is critical for the future 
development of stable high-density integrated circuits.  
Fourier's law of heat conduction is an empirical relationship stating that the flow of heat is linearly related to an applied temperature gradient 
via a geometry independent, but material dependent, thermal conductivity.  

Although Fourier's law accurately describes heat transport in macroscopic samples, at the nanoscale heat is carried by quantum excitations 
(e.g., electrons, phonons, etc.) which are generally strongly influenced by the microscopic details of a system.
For instance, violations of Fourier's law have been observed in graphene nanoribbons, where the system could be tuned between the ballistic phonon regime 
and the diffusive regime by altering the edge state disorder \cite{bae2013ballistic}.  
Violations in carbon nanotubes have also been observed \cite{PhysRevLett.101.075903}.

Investigations into the origin of Fourier's law generally focus on ballistic phonon heat transport. 
However, the electronic heat current can dominate in a variety of systems (e.g., metals, conjugated molecule heterojunctions, etc).  
Unlike phonons, only electrons in the vicinity of a contact's Fermi energy can flow, meaning that wave interference effects play an important role 
in thermal conduction \cite{Bergfield2013demon,bergfield2015tunable}. In addition, the lattice (phonon) and electronic temperatures generally differ
for systems without strong electron-phonon coupling.  In this article, 
we investigate the onset of Fourier's law in the electronic temperature distribution where quantum effects cause the maximal deviations 
from classical predictions.

Previously, Dubi and DiVentra showed that Fourier's law for the electronic temperature could be recovered from a quantum description via two mechanisms: 
dephasing and disorder\cite{dubi2009fourier, dubi2009reconstructing}.  Although valid for some model systems, 
these mechanisms cannot provide a general framework to understand the emergence of Fourier's Law in quantum electron systems. 
The principal shortcoming of these mechanisms, when applied to real nanostructures, is that
the magnitude of dephasing or disorder required to recover Fourier's relation 
is so strong that the covalent bonding of the system would be disrupted,\cite{Bergfield2013demon} effectively disintegrating any real material.

In this work, we utilize a state-of-the-art nonequilibrium quantum description of heat transport to investigate the onset of Fourier's law in a 
nanoscale device.  
Using a non-invasive probe theory \cite{stafford2016local,stafford2017local} in which the spatial resolution of the temperature measurement is limited by 
fundamental thermodynamic relationships rather than by the stucture and composition of the probe, we find that Fourier's law emerges in the limit
where many quantum states contribute to the heat transport.  That is, when the energy-level spacing of the quantum states of the system is small compared to
the coupling of the system to the source and drain reservoirs, so that 
the density of states of a system becomes smooth.
Finally, we apply a thermal resistor network analysis to the simulated temperature profiles and 
observe the emergence of a geometry-independent thermal conductivity.

\begin{figure*}[tb]
	\centering
	\subfloat[Contact type I - Classical]{\includegraphics[width=0.35\linewidth]{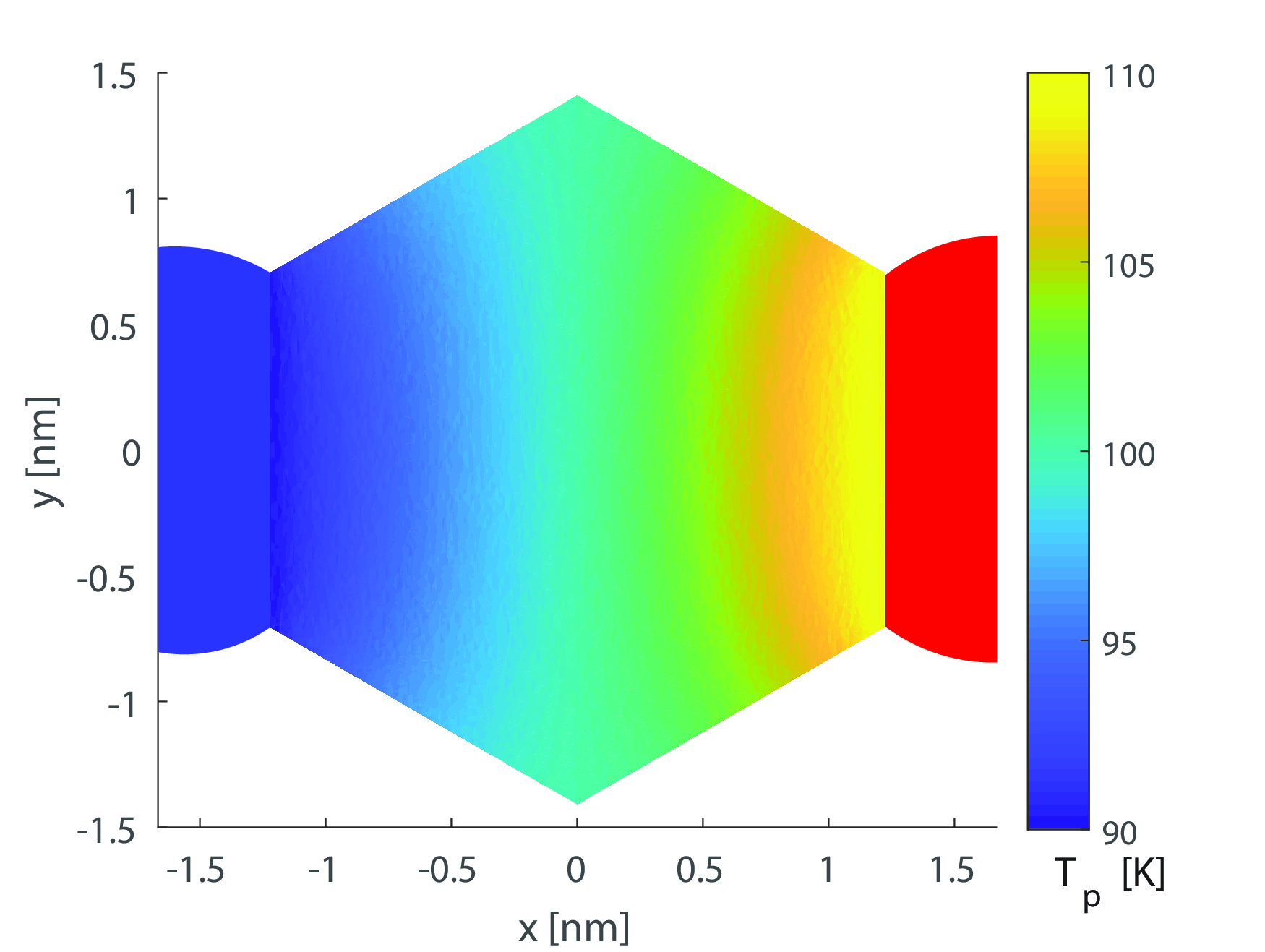}\label{fig:ClassicalType1}}
	\subfloat[Contact type I - Quantum]{\includegraphics[width=0.35\linewidth]{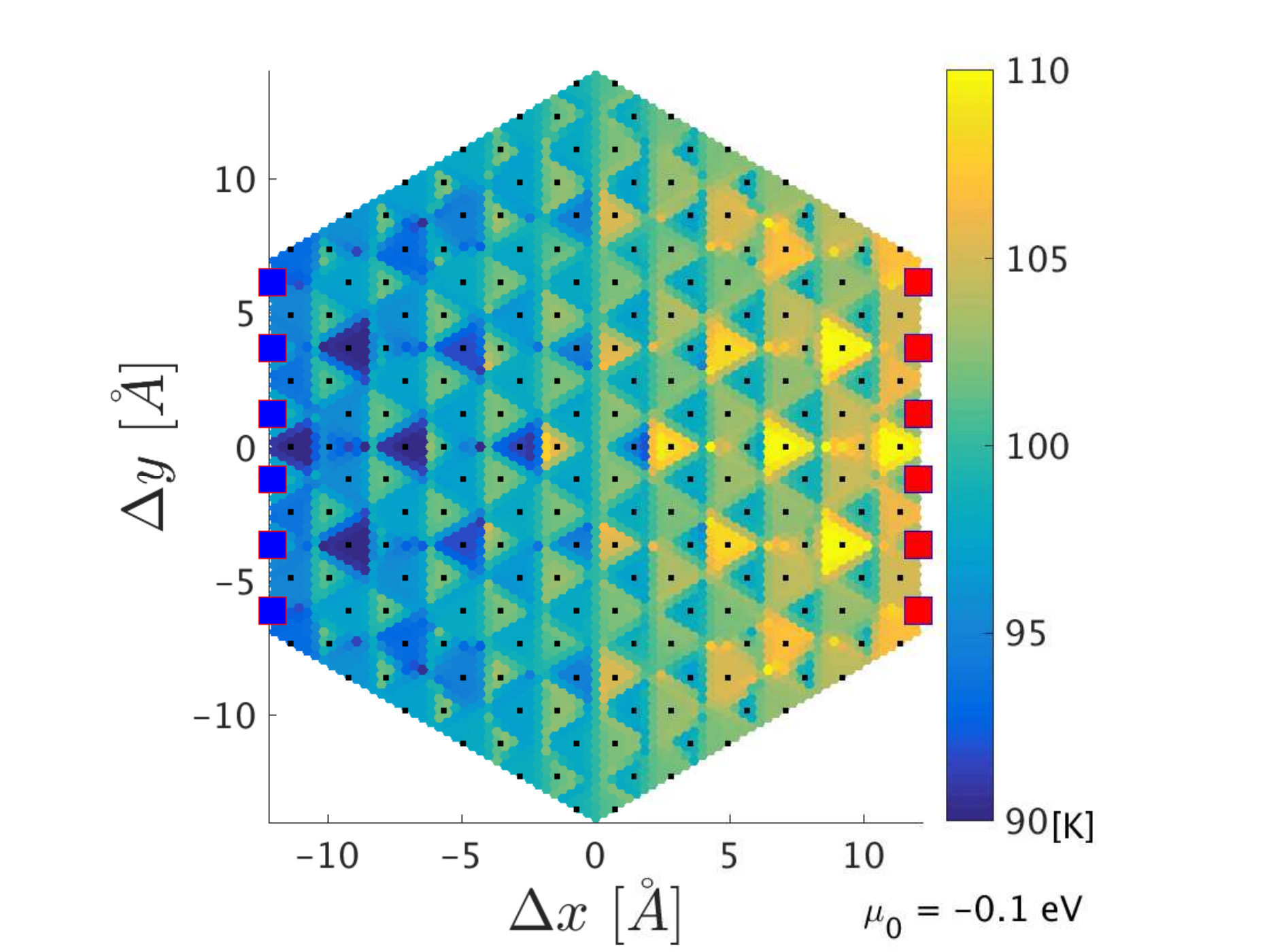}\label{fig:mu01_C1_gamma3}}%
	\subfloat[Contact type I  - Quantum]{\includegraphics[width=0.35\linewidth]{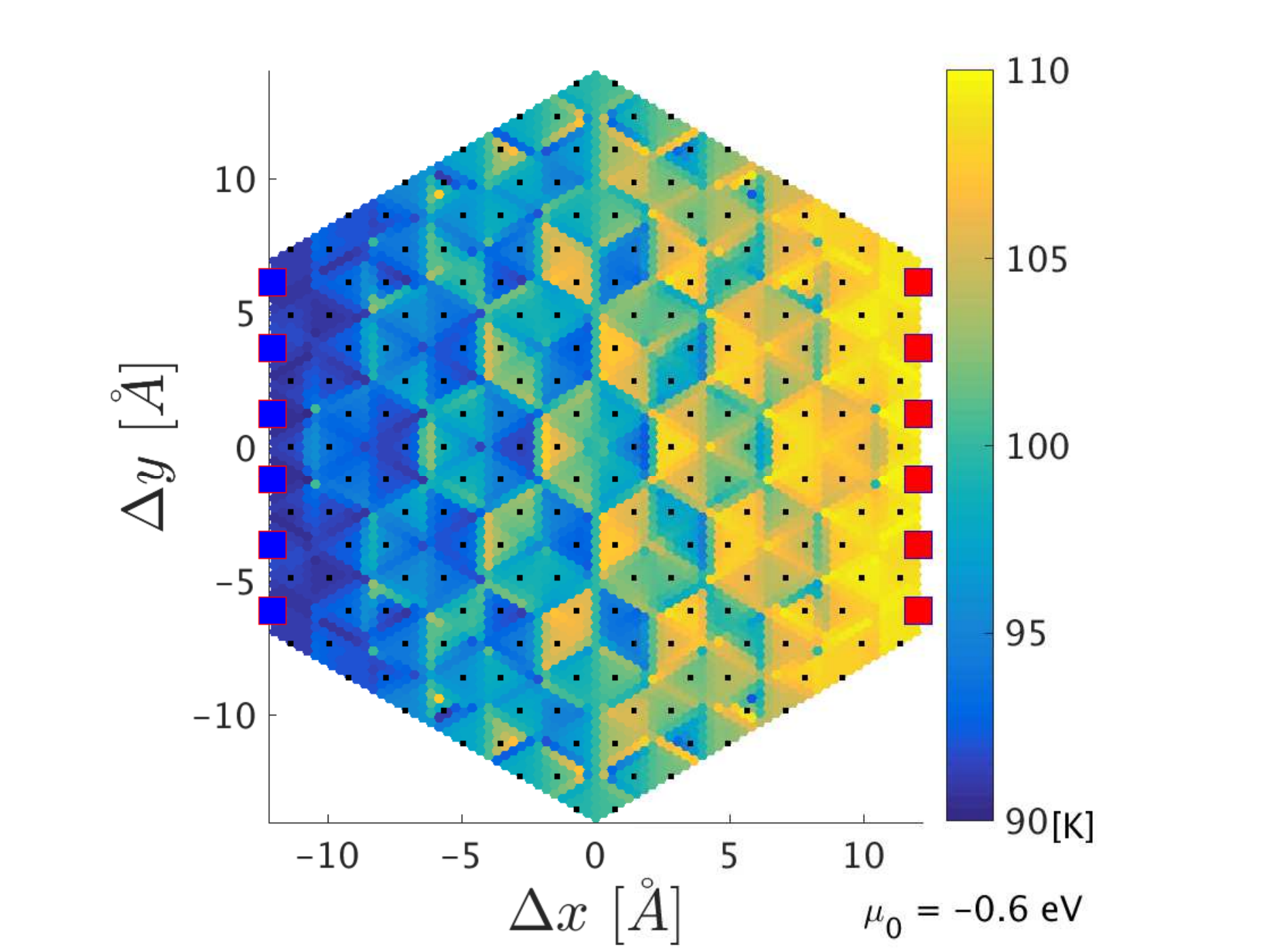}\label{fig:mu06_C1_gamma3}}%
	\\
	\vspace{-0.35cm}
	\subfloat[Contact type II - Classical]{\includegraphics[width=0.35\linewidth]{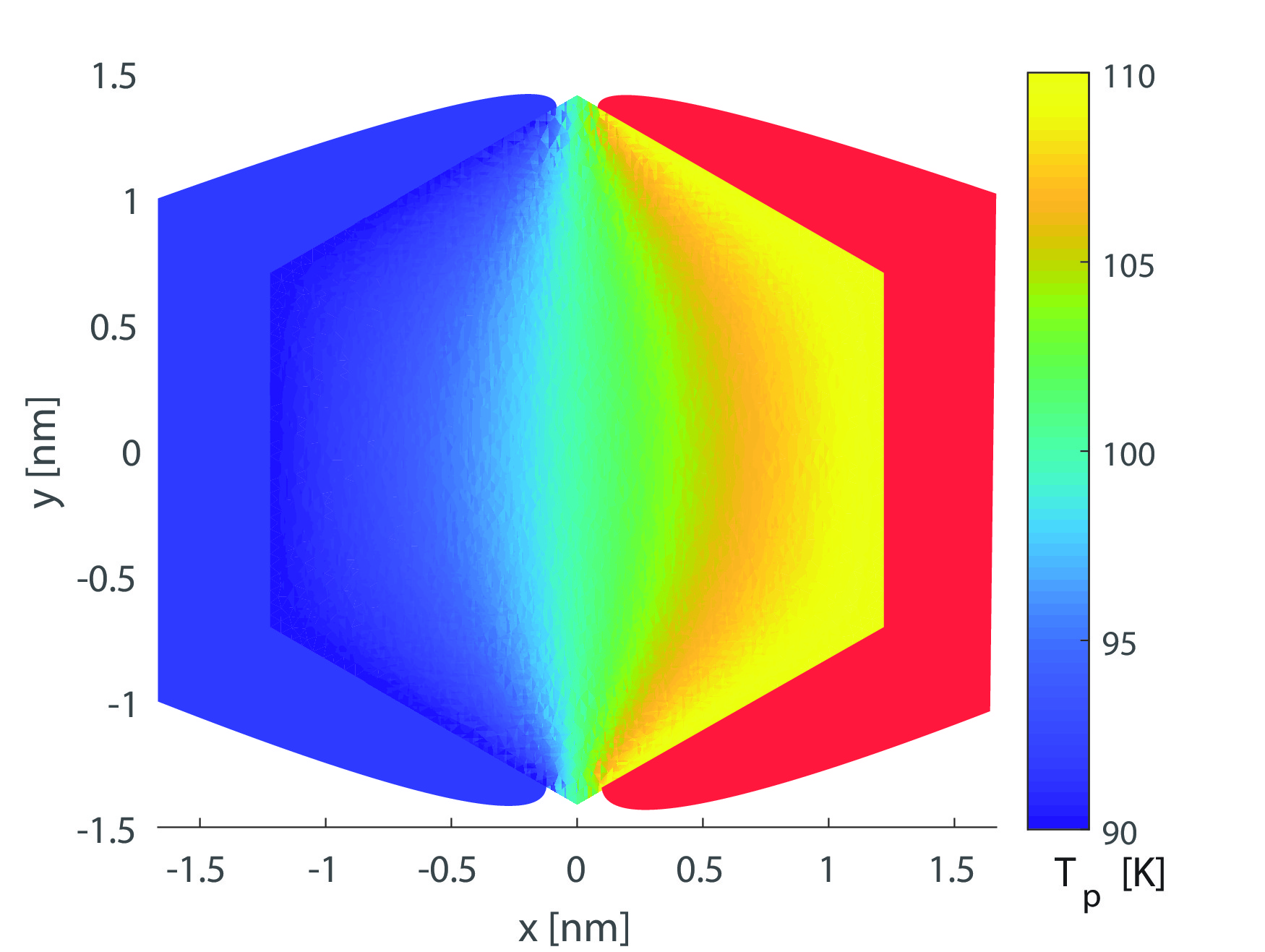}\label{fig:ClassicalType2}}
	\subfloat[Contact type II - Quantum]{\includegraphics[width=0.35\linewidth]{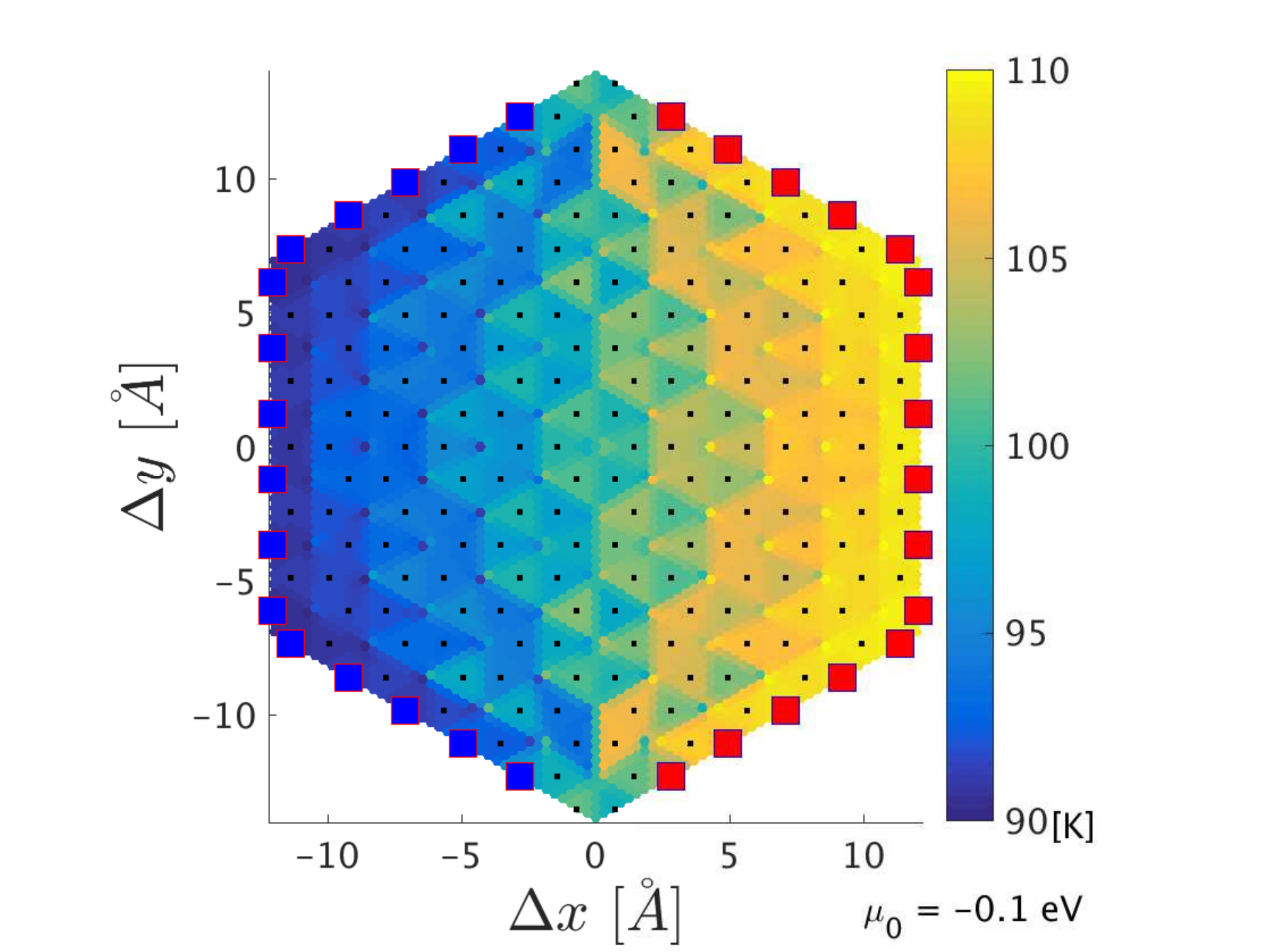}\label{fig:mu01_C5_gamma3}}%
	\subfloat[Contact type II -  Quantum]{\includegraphics[width=0.35\linewidth]{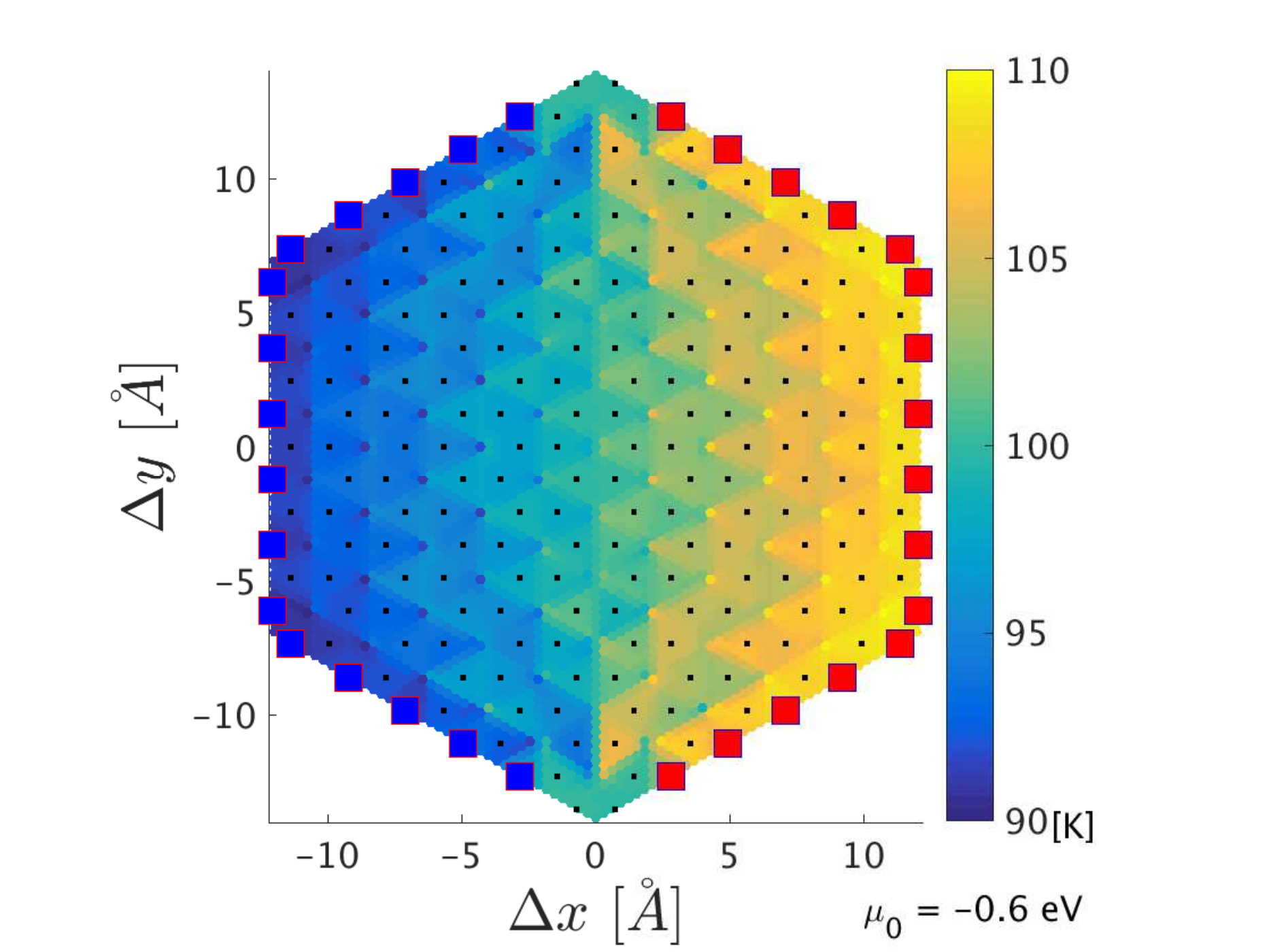}\label{fig:mu06_C5_gamma3}}%
	\vspace{-0.2cm}
	\caption{Classical (panels a,d) and quantum temperature profiles of a graphene flake under thermal bias for two contact geometries. 
		The hot electrode (red) is held at 110K and the cold electrode (dark blue) is held at 90K, where red and blue squares indicate the carbon atoms covalently 
		bonded to the hot and cold electrodes, respectively.
		In contact type I (upper panels), only the left and right edges of the flake couple to the electrodes, while in contact type II, the coupling to the 
		electrodes wraps around three edges each, leading to three times stronger coupling to the electrodes.
		The quantum calculations are at Fermi energies $\mu_0=-0.1$eV (b,e) and $-0.6$eV (c,f), relative to the Dirac point.    
		The quantum temperature distributions for contact type I exhibit strong oscillations that depend sensitively on $\mu_0$, while for contact type II, the
		temperature distributions resemble pixelated versions of the classical distribution.
	}
	\label{fig:temperature_profile}%
\end{figure*}

\section{Theory of local temperature measurement}

Fourier's law for the heat current density $J_q = -\kappa \nabla T$
establishes a local linear relationship between an applied temperature gradient $\nabla T$ 
and the heat flow, and is generally accurate for macroscopic, dissipative systems.  
In quantum systems, the local temperature $T({\bf x})$ must be thought of as the result of a local {\em measurement}, 
and can vary due to quantum interference effects,\cite{bergfield2015tunable,Bergfield2013demon} quantum chaos \cite{lepri1997heat}, 
disorder\cite{dubi2009reconstructing}, and dephasing\cite{dubi2009fourier} of the heat carriers in the sample.

The local temperature distribution of a nonequilibrium quantum system is defined by introducing a floating thermoelectric probe 
\cite{Bergfield2013demon,Meair14,bergfield2015tunable,shastry2016temperature}.  
The probe exchanges charge and heat with the system via a local coupling until it reaches equilibrium with
the system:
\begin{equation}
I_p^{(\nu)} =0, \; \nu=0,1,
\label{eq:def_probe}
\end{equation}
where $-eI^{(0)}_p$ and $I^{(1)}_p$ are the electric current and heat current, respectively, flowing into the probe.
The probe is then in local equilibrium with a quantum system which is itself out of equilibrium.  

In the linear-response regime, for a thermal bias applied between electrodes 1 and 2, forming an open electric circuit, the heat current into electrode $\alpha$ is given by
\begin{equation}
	I^{(1)}_\alpha = \sum_\beta \tilde{\kappa}_{\alpha\beta} (T_\beta - T_\alpha),
	\label{eq:heat_current}
\end{equation}
where $\alpha$ and $\beta$ label one of the three electrodes (1, 2, or the probe).  Solving this set of linear equations, we arrive at the local temperature 
distribution \cite{Bergfield2013demon}
\begin{equation}
T_p(x,y)=\frac{\tilde{\kappa}_{p1}(x,y) T_1 + \tilde{\kappa}_{p2}(x,y) T_2 + \kappa_{p0} T_0}{
\tilde{\kappa}_{p1}(x,y) + \tilde{\kappa}_{p2}(x,y) +  \kappa_{p0}}.
\label{eq:Tp}
\end{equation}
Here $\tilde{\kappa}_{p\beta}(x,y)$ is the 
position-dependent
thermal conductance between electrode $\beta$ and the probe, and $\kappa_{p0}$ is the thermal
coupling of the probe to the ambient environment at temperature $T_0$.

In the absence of an external magnetic field, 
the 
effective two-terminal thermal conductances 
are given by \cite{Bergfield2013demon}
\begin{eqnarray}
\tilde{\kappa}_{\alpha\beta} &=& \frac{1}{T}\left[\Lfun{2}{\alpha\beta} 
-\frac{\left[\Lfun{1}{\alpha\beta}\right]^2}{\tilde{\gcal L}_{\alpha\beta}^{(0)}} \right. \nonumber \\
& -&  \left.  {\gcal L}^{(0)} \!
\left(
\frac{\Lfun{1}{\alpha\gamma}\Lfun{1}{\alpha\beta}}{\Lfun{0}{\alpha\gamma}\Lfun{0}{\alpha\beta}}
+\frac{\Lfun{1}{\gamma\beta}\Lfun{1}{\alpha\beta}}{\Lfun{0}{\gamma\beta}\Lfun{0}{\alpha\beta}}
-\frac{\Lfun{1}{\alpha\gamma}\Lfun{1}{\gamma\beta}}{\Lfun{0}{\alpha\gamma}\Lfun{0}{\gamma\beta}}
\right)
\right], 
\label{eq:kappatilde}
\end{eqnarray}
where $\Lfun{\nu}{\alpha\beta}$ in an Onsager linear 
response 
function 
\cite{Onsager31},
\begin{equation}
\tilde{\gcal L}_{\alpha\beta}^{(0)}=    \Lfun{0}{\alpha\beta}+
\frac{\Lfun{0}{\alpha\gamma}\Lfun{0}{\gamma\beta}}{\Lfun{0}{\alpha\gamma}+\Lfun{0}{\gamma\beta}},
\label{eq:three_term_L0}
\end{equation}
and 
\begin{equation}
\frac{1}{{\gcal L}^{(0)}} = \frac{1}{\Lfun{0}{12}} + \frac{1}{\Lfun{0}{13}} + \frac{1}{\Lfun{0}{23}}.
\label{eq:L0_series}
\end{equation}

Following the methods of Refs.~\onlinecite{Sivan86,Bergfield09b,Bergfield10},
the linear-response coefficients may be calculated in the elastic cotunneling regime as 
\begin{equation}
\Lfun{\nu}{\alpha\beta} (\mu) = \frac{1}{h} \int dE \left(E-\mu \right)^\nu \left(-\frac{\partial f}{\partial E}\right) T_{\alpha\beta}(E),
\label{eq:Lnu}
\end{equation}
where $f(E)$ is the equilibrium Fermi-Dirac distribution, and $T_{\alpha\beta}(E)$ is the transmission probability from contact $\beta$ to contact $\alpha$
for an electron of energy $E$, which may be found using the usual nonequilibrium Green's function (NEGF) methods. 
The details of our computational methods may be found in the Supporting Information.

In the simulations discussed below, we consider an ideal broad-band probe with perfect spatial resolution
coupled weakly to the system.\cite{stafford2016local,stafford2017local}  
Furthermore, we assume $\kappa_{p0} \ll \tilde{\kappa}_{p1}, \, \tilde{\kappa}_{p2}$, so that we can unambiguously determine the fundamental value of the 
local temperature in the nonequilibrium system.  Any actual scanning probe won't achieve this resolution; instead a convolution between the intrinsic profile 
and the probe's resolution will be measured.  The advantage of considering a probe in this limit is that we can investigate the onset of Fourier's law without the complications introduced by the probe's apex wavefunction geometry.

\section{Results for $T_p(x,y)$ in graphene nanojunctions}

We investigate heat transport and temperature distributions in a 
graphene flake coupled to two macroscopic metal electrodes under a thermal bias. 
The electrodes are covalently bonded to the edges of the graphene flake. 
See Supporting Information for details of the model.

\subsection{Emergence of Fourier's law}
The classical temperature distribution for a graphene flake with two different contact geometries 
is shown in panels a and d of Fig.~\ref{fig:temperature_profile}. The behavior predicted by Fourier's law is clearly visible in the 
characteristic linear temperature gradient across the sample from the hot to the cold electrode.  
This behavior is to be contrasted with the temperature distributions calculated using quantum heat transport theory, shown in panels b, c, e, and f.
Figs.~\ref{fig:temperature_profile}b, c show the electron temperature distributions for two different values of the Fermi energy
($\mu_0=-0.1eV, -0.6eV$ relative to the Dirac point) for contact type I, where the hot and cold electrodes are covalently
bonded to the right and left edges of the graphene flake at the sites indicated by red squares.  
The temperature exhibits large quantum oscillations\cite{bergfield2015tunable} that depend sensitively on the Fermi energy $\mu_0$, obscuring any possible
resemblance to the classical temperature distribution shown in Fig.\ \ref{fig:temperature_profile}a.  The electron temperature
distributions for contact type II are shown for the same two values of $\mu_0$ in Figs.~\ref{fig:temperature_profile}d, e.  In this case, although there are
atomistic deviations from Fourier's law, nonetheless the resemblance to the classical distribution shown in Fig.\ \ref{fig:temperature_profile}d is
unmistakable, and there is not a strong dependence on $\mu_0$.

\begin{figure}[tb]
	\centering
	\includegraphics[width=3in]{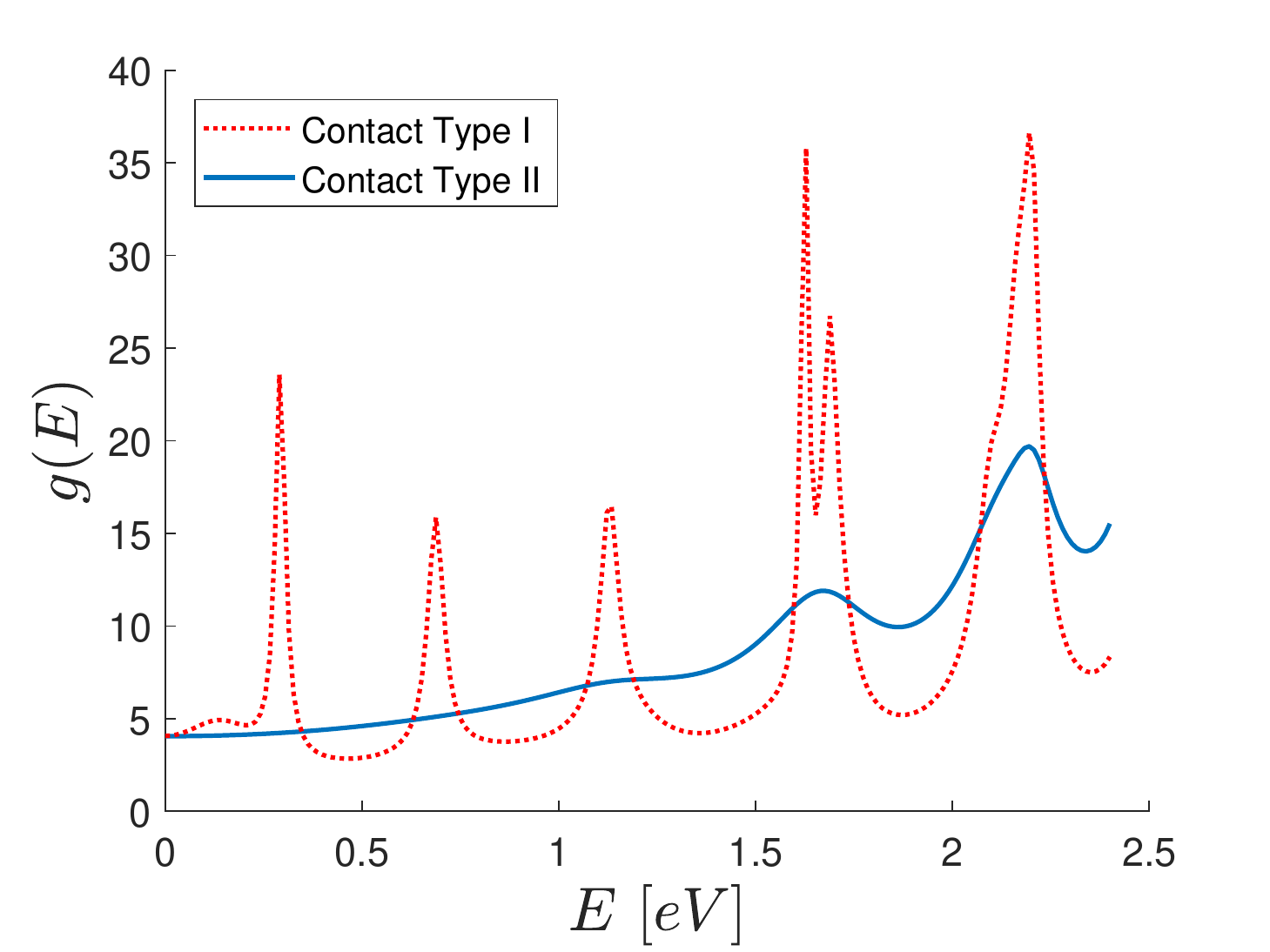}
	\caption{The calculated density of states (DOS) $g(E)$ of a graphene flake junction for two different contact geometries, defined in 
		Fig.\ \ref{fig:temperature_profile}.
		The DOS for contact type I exhibits a sequence of sharp peaks corresponding to the energies of individual energy eigenstates 
		[or manifolds of (nearly) degenerate eigenstates] of the flake, broadened by coupling to the electrodes.  Contact type II, for which  
		the broadening is three times as large, exhibits a smooth, nearly featureless DOS for $E<1.2$eV.
	}
	\label{fig:DOS}
\end{figure}

The different nature of thermal transport for contact types I and II can be understood by considering the density of states (DOS) $g(E)$ of the system, 
shown in Fig.~\ref{fig:DOS}. For contact type I, $g(E)$ exhibits a sequence of well 
defined peaks, corresponding to the energy eigenfunctions of the graphene flake broadened by coupling to the leads. 
In constrast, contact type II, where the broadening is three times as large, has a smooth, almost featureless DOS for $E<1.2$eV.  
A sharply-peaked DOS indicates that the system is in the {\em resonant-tunneling regime} where
thermal transport is controlled by the wavefunction of a single resonant state [or a few (nearly) degenerate states], while a smooth DOS indicates that
many quantum states contribute to thermal transport, so that quantum oscillations tend to average out.
We find that a necessary and sufficient condition to recover Fourier's law is that many (nondegenerate) quantum states contribute with comparable strength
to the thermal transport.  When transport occurs in or near the resonant-tunneling regime, on the other hand,
there is no classical limit for the temperature distribution.  

\begin{figure}[htb]
	\centering
	\subfloat[Contact type I]{\includegraphics[width=\linewidth]{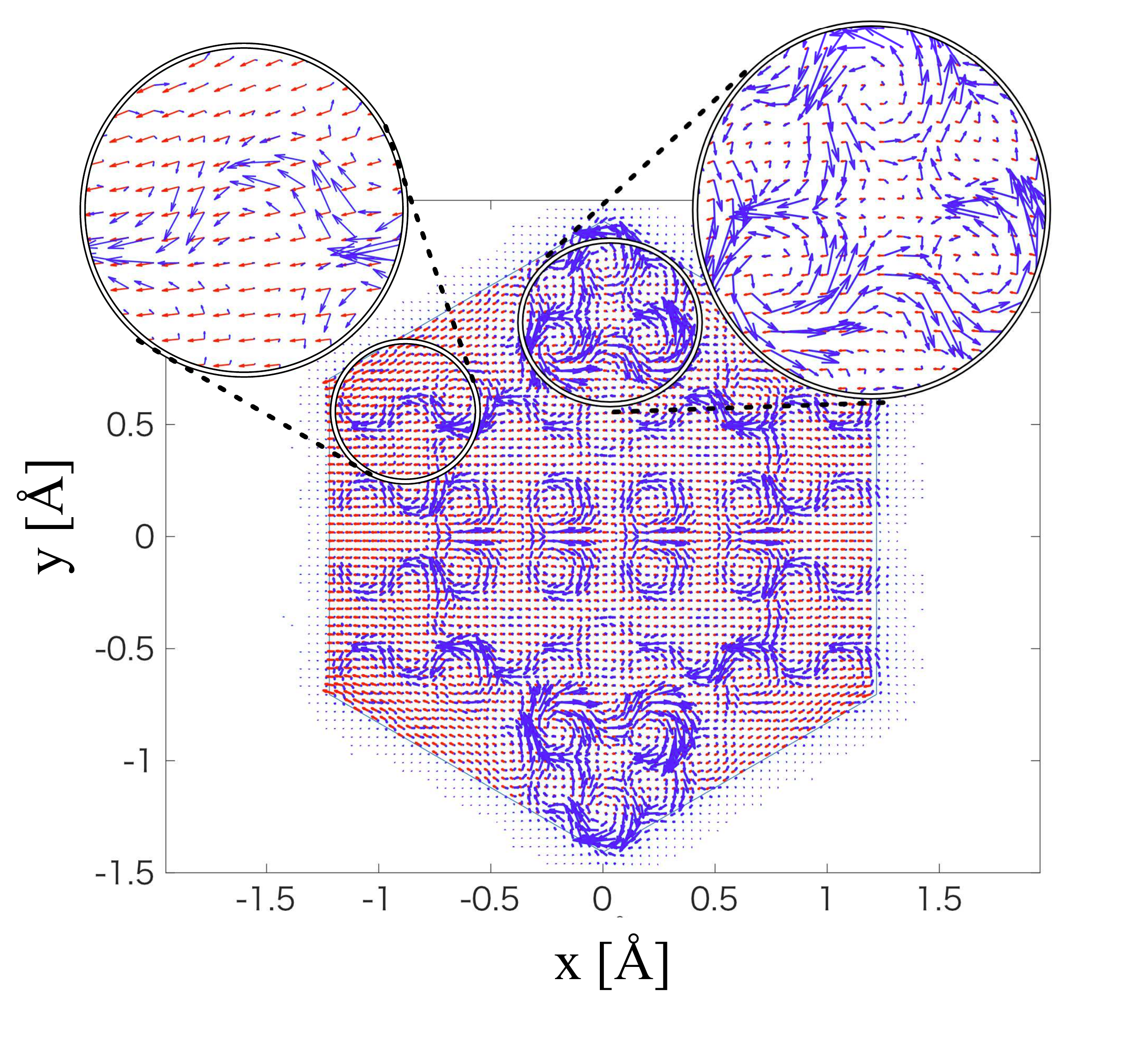}}
	\\
	\vspace{-.4cm}
	\subfloat[Contact type II]{\includegraphics[width=\linewidth]{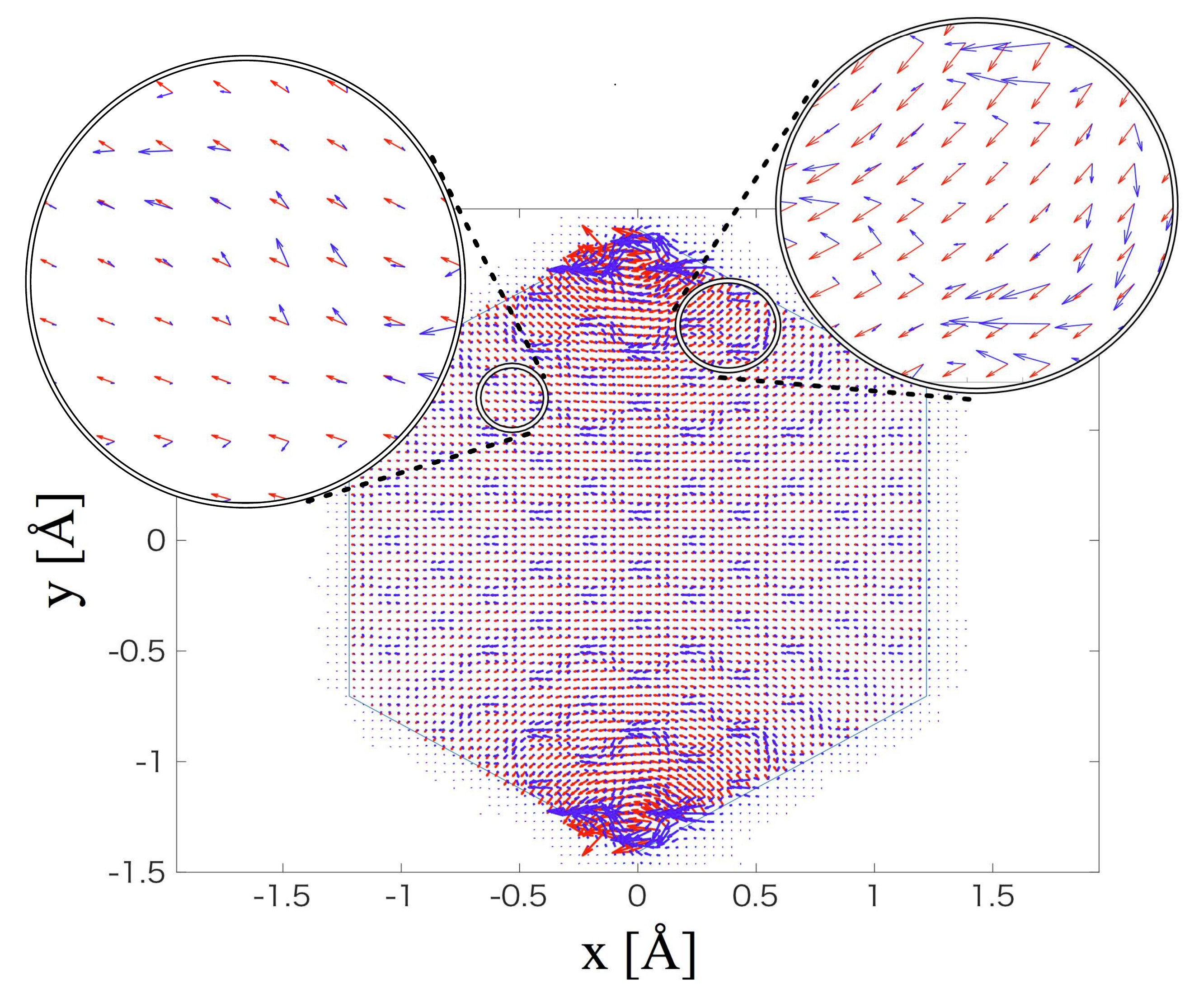}}	
	\caption{Top panel: The heat current density $\vec{J}_Q$ for contact type I at $\mu_0 = -2.2$eV is calculated using classical, and quantum transport theories, indicated with the red and blue arrows, respectively.  Bottom panel: $\vec{J}_Q$ for contact type II at $\mu_0 = -0.1$eV 
		calculated using classical, and quantum transport theories. As highlighted by the swirling blue arrows, the heat current profile of the junction shown in the top panel is highly non-classical, while the heat transport profile of the bottom panel's junction is better represented by a classical description. 
	} 
	\label{fig:heat_flux}
\end{figure}


We note that for nanostructures amenable to simulation (a few hundred atoms or less),
a very large coupling to the electrodes is necessary to push the system out of the resonant-tunneling regime, and we speculate that this may be the
reason why attempts to study the quantum to classical crossover in electron thermal transport via simulation have so far proven problematic.
Thermal transport experiments
are routinely conducted with much larger quantum systems, however, where this condition is well satisfied.

A direct test of Fourier's law involves not only the temperature distribution but also the heat current density $J_q$, which may be calculated using
NEGF methods (see Supporting Information).
The simulated heat flow patterns are shown in Fig.~\ref{fig:heat_flux} for both classical and quantum thermal transport in both contact geometries.
The quantum heat flow in contact type I bears little relation to the classical flow, but instead exhibits vortices and fine structure that is strongly
energy dependent, similar to the 
local charge current structure \cite{solomon2010exploring}. 
In contrast, the quantum heat flow in contact type II is nearly classical, except that it is concentrated along the C---C bonds, which serve as conducting
channels.  The heat flow patterns shown in Fig.\ \ref{fig:heat_flux} confirm that the crossover to the classical thermal transport regime requires 
many quantum states of the graphene flake to contribute comparably (smooth DOS).

\subsection{Thermal resistor network model}

\begin{figure*}[htb]%
		\centering
\includegraphics[width=.45\linewidth]{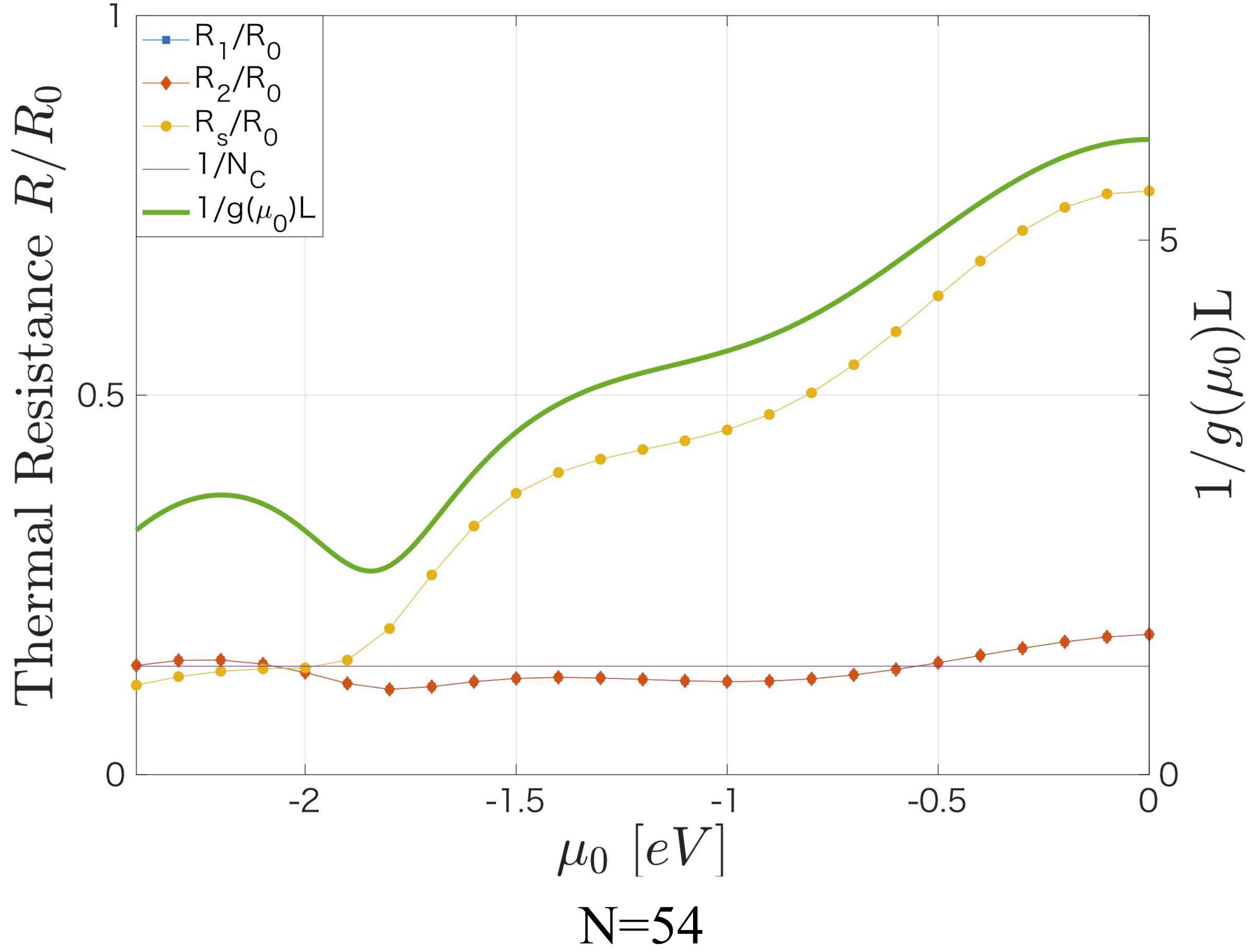}\quad\quad
\includegraphics[width=.45\linewidth]{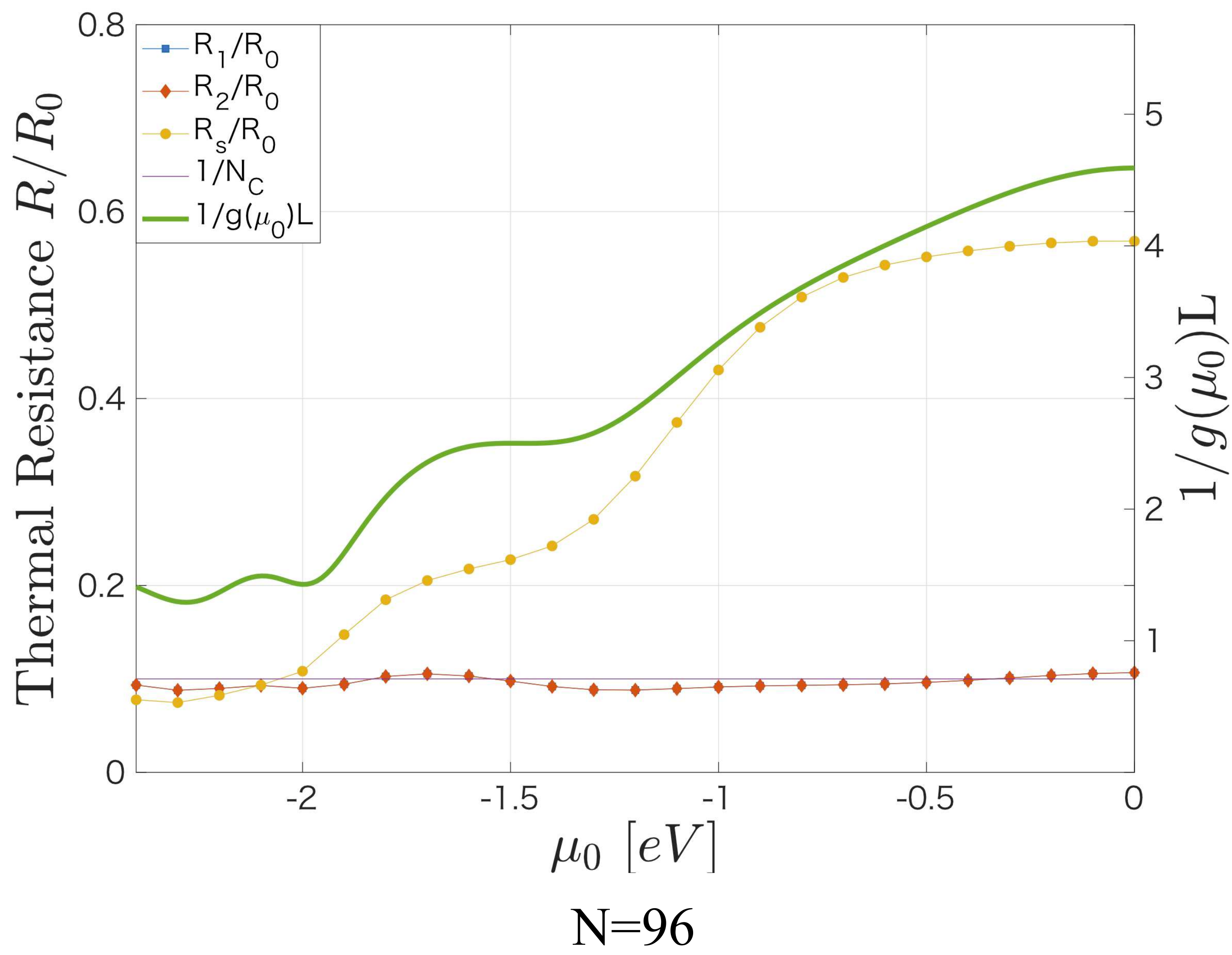}\\
	\includegraphics[width=.45\linewidth]{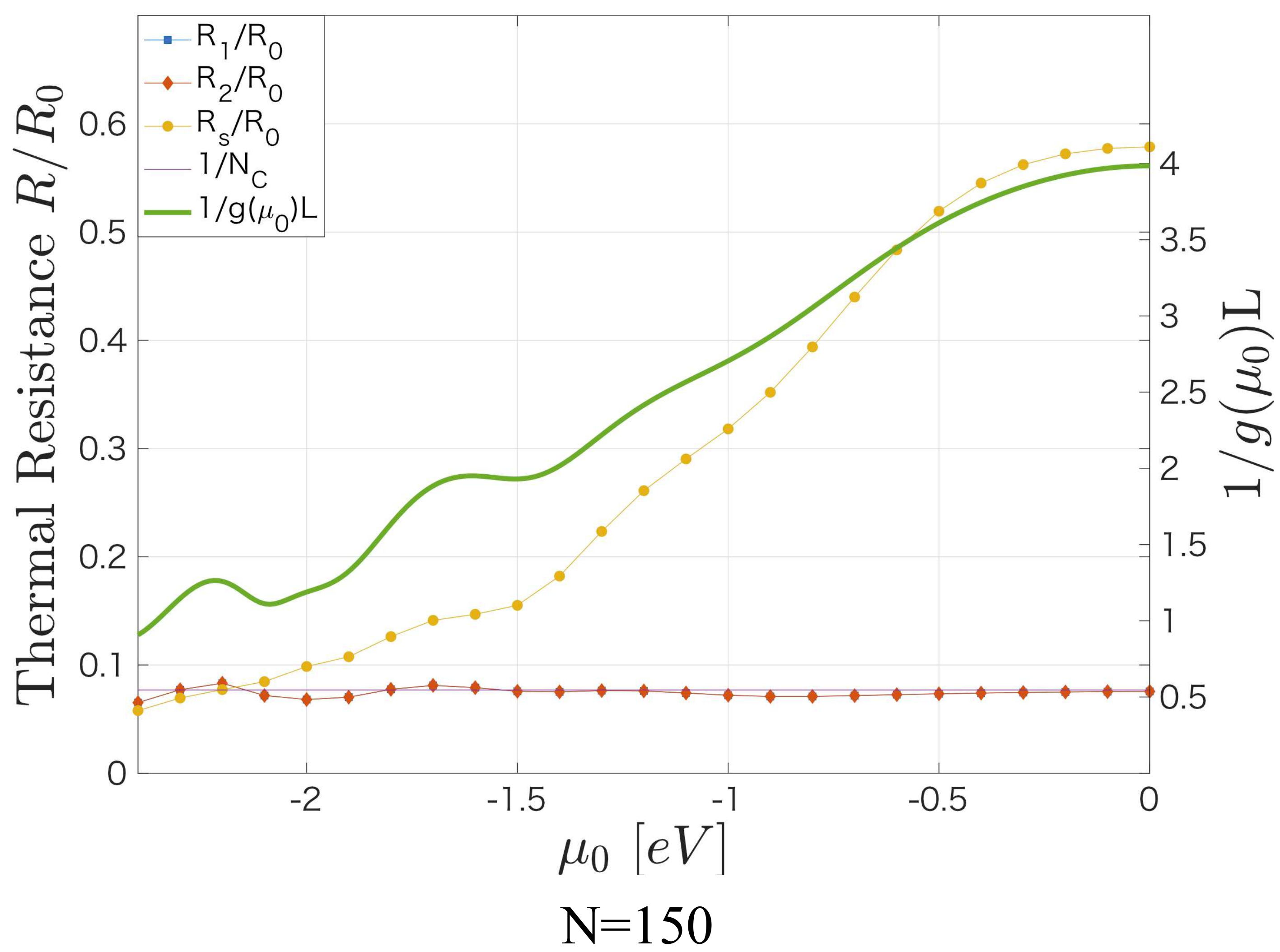}\quad\quad
\includegraphics[width=.45\linewidth]{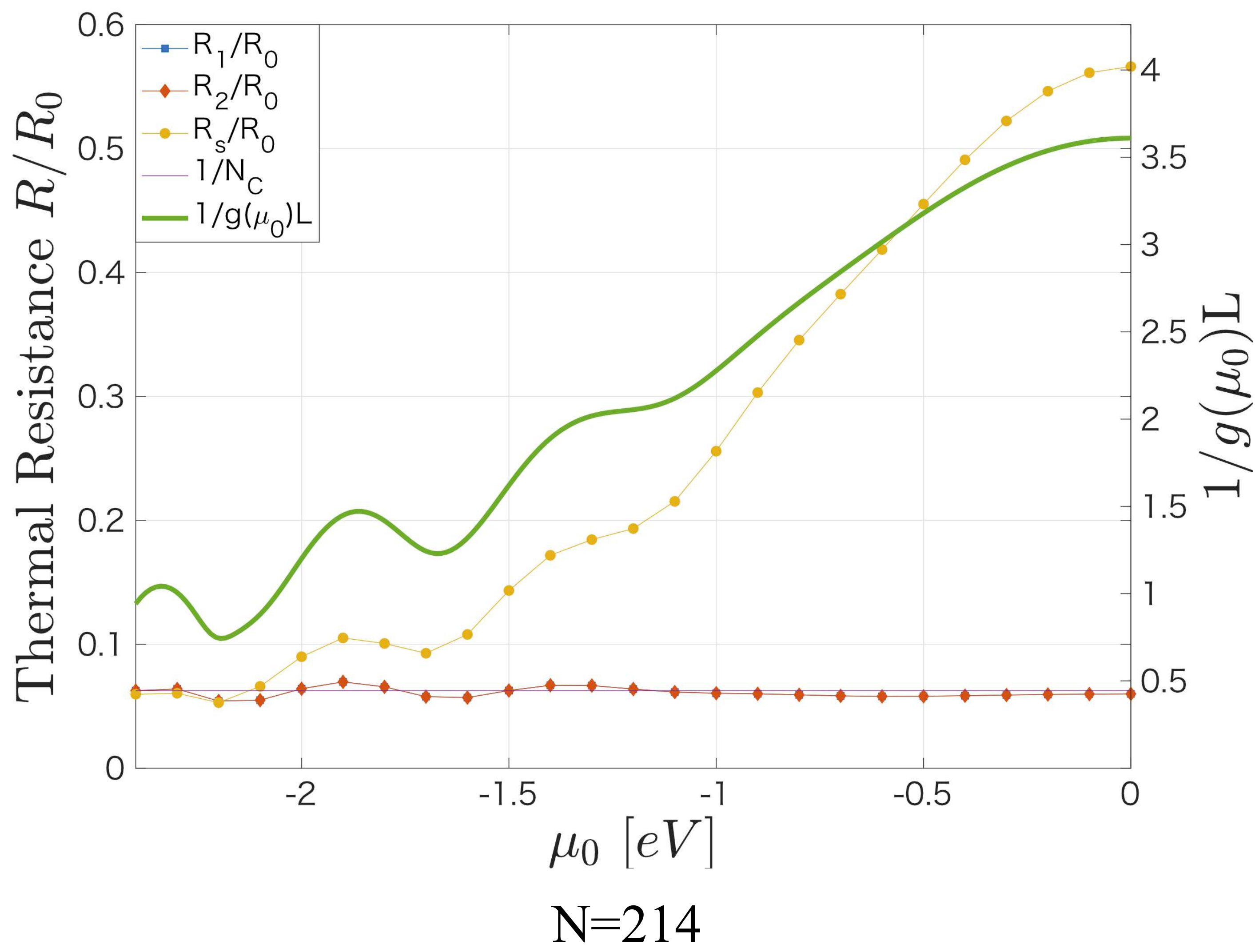}
	\caption{Thermal resistance values as a function of Fermi energy $\mu_0$ 
for four different sized hexagonal graphene flake junctions with contact type II, where $N$ is the number of atoms in the flake. 
$R_s$ is the sample thermal resistance and $R_1$ and $R_2$ are the contact thermal resistances, defined in Eqs.\ (\ref{eq:Rs})--(\ref{eq:R2}), 
respectively.
The contact resistances are nearly universal in this transport regime: $R_1,\,R_2\approx R_0/N_c$, where $R_0$ is the thermal resistance quantum and 
$N_c$ is the number of atoms bonded to each contact.
The sample thermal resistance $R_s$ is inversely correlated with the density of states per unit area times the sample length,
$g(\mu_0)L$, as expected based on semiclassical
Boltzmann transport theory.
}
	\label{fig:R_vs_size}
\end{figure*}

The thermal conductivity $\kappa$ in Fourier's law is  material dependent but dimensionally independent.  
In the regime of quantum transport \cite{Datta95}, linear response theory instead treats the thermal conductance (also traditionally denoted by the symbol
$\kappa$), which depends in detail on the dimensions and structure of the conductor.
In order to investigate the cross-over between these regimes, we develop a thermal circuit model and apply 
it to the temperature profiles calculated using our theory.  

The temperature probe acts as a third terminal in the thermoelectric circuit, and affects the thermal conductance between the hot and cold electrodes.  
Starting from Eq.~(\ref{eq:heat_current}), the heat current flowing into electrode $1$ may be expressed as
\begin{equation}
I_1^{(1)}= \tilde{\tilde{\kappa}}_{12} (T_2-T_1),
\label{eq:def_Rth}
\end{equation}
where the thermal conductance between source and drain in the presence of the thermal probe is 
\begin{equation}
\tilde{\tilde{\kappa}}_{12}=\tilde{\kappa}_{12} + \frac{\tilde{\kappa}_{p1} \tilde{\kappa}_{p2}}{\tilde{\kappa}_{p1}+\tilde{\kappa}_{p2}}.
\label{eq:kappa12}
\end{equation}

The thermal resistance of the junction may be written as
\begin{equation}
R_{\rm th} \equiv \tilde{\tilde{\kappa}}_{12}^{-1} = R_s + R_1 + R_2,
\label{eq:R_network}
\end{equation}
where $R_s$ is the ``intrinsic'' thermal resistance of the system, and
$R_1$ and $R_2$ are thermal contact resistances associated with the interfaces between electrodes 1 and 2, respectively, and the quantum
system.

The individual resistances in the network are defined as follows:
\begin{eqnarray}
R_s & = & \frac{|T_{1s} - T_{2s}|}{|I_1^{(1)}|}, 
\label{eq:Rs}
\\
R_1 & = & \frac{|T_{1} - T_{1s}|}{|I_1^{(1)}|}, 
\label{eq:R1}
\\
R_2 & = & \frac{|T_{2s} - T_{2}|}{|I_1^{(1)}|}, 
\label{eq:R2}
\end{eqnarray}
where $T_{\alpha s}$ is the temperature averaged over the atoms bonded to electrode $\alpha$.

The contact resistances $R_1$, $R_2$, and sample thermal resistance $R_s$
are shown for four different sized graphene flakes 
with contact type II as a function of Fermi energy in Fig.~\ref{fig:R_vs_size}.  
Here $N$ is the number of atoms in the hexagonal flake.
The resistances are normalized by the quantum of thermal resistance $R_{0} = {3h}/{\pi^2 k_B^2 T_0}\simeq 1.2\times 10^{9}K/W$ at $T_0$=100K.  
For these junctions, the contact resistances exhibit nearly universal behavior
\begin{equation}
R_1, \, R_2 \approx R_0/N_c,
\label{eq:R_1,2}
\end{equation}
where $N_c$ is the number of atoms bonded to each contact, with only small deviations that decrease in amplitude with increasing flake size.

To study the crossover to the classical transport regime, it is useful to compare the sample thermal resistance $R_s$ to the classical result derived from
a two-dimensional Boltzmann equation in the relaxation-time approximation
\begin{equation}
R_{\rm cl} = \frac{2 R_0}{h L g(\mu_0) v_F},
\label{eq:R_cl}
\end{equation}
where $g(E)$ is the density of states per unit area of the graphene flake, $v_F$ is the Fermi velocity, and we have set the scattering time
$\tau=L/v_F$ for these ballistic conductors, where $L$ is the distance between the source and drain electrodes.
Note that these hexagonal flakes have equal width and length, so the geometric factor in 
$R_{\rm cl}$ is unity.
Eq.\ (\ref{eq:R_cl}) implies that near the Dirac point in graphene, where $v_F\approx \mbox{const.}$, $R_{\rm cl} \propto 1/g(\mu_0)L$.
Fig.\ \ref{fig:R_vs_size} shows that indeed the variations of $R_s$ with Fermi energy are correlated with the variations of $1/g(\mu_0)L$ for various
flake sizes, confirming the classical nature of transport in junctions with contact type II.
An improved fit might be obtained by including the variation of $v_F$ with $\mu_0$, which is important far from the Dirac point.

Although the temperature distribution can approach the classical 
limit in some cases via coarse graining \cite{Bergfield2013demon}, the thermal resistor network 
model is found to be quantitatively consistent with Fourier's law only for the nearly classical transport regime, 
where multiple resonances contribute to the transport.
In the quantum transport regime, where individual resonances are important, the contact resistances are not universal, 
but exhibit large oscillations as well, 
making the identification of a ``sample thermal resistance'' problematic.  

%
%
%

\section{Conclusions}

Thermal transport in quantum electron systems was investigated, and the crossover from the quantum transport regime to the classical
transport regime, where Fourier's law holds sway, was analyzed.  In the quantum regime of electron thermal transport, the local temperature
distributions exhibit large oscillations due to quantum interference\cite{DiVentra09,Bergfield2013demon,Meair14,bergfield2015tunable} 
(see Fig.\ \ref{fig:temperature_profile}b,c),
and the heat flow pattern exhibits vortices and other nonclassical features (see Fig.\ \ref{fig:heat_flux}b),
while in the classical regime, the heat flow is laminar (see Fig.\ \ref{fig:heat_flux}d) 
and the temperature drops monotonically from the hot to the cold electrode (see Fig.\ \ref{fig:temperature_profile}e,f). 

A satisfactory understanding of the quantum to classical crossover in electron thermal transport has been lacking for a number of years.  Perhaps the most
promising explanation advanced early on\cite{Dubi09b} was in terms of dephasing of the electron waves:  for sufficiently large inelastic scattering in the
system, the electron thermal transport becomes classical.  However, many nanostructures of interest for technology, such as graphene, have very weak
inelastic scattering \cite{Hwang08}, and the origin of Fourier's law cannot be explained in such systems by this mechanism.

In this article, it was shown that a sufficient condition for a quantum electron system to cross over into the classical thermal transport regime is
for the broadening of the energy levels of the system to exceed their separation, so that the DOS 
becomes smooth.  
In this limit, the transport involves contributions from multiple resonances above and below the Fermi level, so that interference effects average out.
This condition is challenging to achieve in simulations, requiring
almost the entire edge of the largest 2D system studied to be covalently bonded to one of the two electrodes (see Fig.\ \ref{fig:temperature_profile}d--f).
For smaller systems, unphysically large
electrode coupling would be required to reach the classical regime.  However, for the larger systems routinely studied in experiments
\cite{Jiamin12}, it may be quite
typical for thermal transport to occur in the classical regime, since the level spacing scales inversely with the system size.

In addition to recovering a nearly classical temperature profile in the limit where the DOS is smooth, it was also shown that the thermal resistance of the
junction could be explained using a thermal resistor network model consistent with Fourier's law in this limit.  The contact thermal resistances were 
found to take on universal quantized values, while the sample thermal resistance was found to be inversely proportional to the DOS per unit area times 
the sample length, as expected based
on semiclassical Boltzmann transport theory (see Fig.\ \ref{fig:R_vs_size}).  In contrast, in the quantum regime, where thermal transport occurs predominantly
via a single energy eigenstate (or a few closely-spaced states near the Fermi level), 
the thermal resistor network model was not found to be useful in analyzing the transport.
In this sense, coarse graining of the temperature distribution due to limited spatial resolution of the probe, which leads in many cases 
\cite{Bergfield2013demon} to a rather classical temperature profile, is not sufficient to explain the onset of Fourier's law, since the underlying
thermal transport remains quantum mechanical.

\begin{acknowledgments}
We acknowledge useful discussions with Brent Cook during the early stages of this project.
J.P.B. was supported by an Illinois State University NFIG grant. 
C.A.S.\ was supported by the U.S.\ Department of Energy (DOE), Office of Science under Award No.\ DE-SC0006699.
\end{acknowledgments}

\section{Appendix}

We utilize a standard nonequilibrium Green's function (NEGF) framework\cite{Datta95,Bergfield09a} 
to describe the quantum transport through a three-terminal junction composed 
of a graphene flake coupled to source and drain electrodes, and a scanning probe.  We focus on transport in the elastic cotunneling regime, where
the linear response coefficients may be calculated from the transmission coefficients $T_{\alpha\beta}(E)$ using 
Eq.\ (\ref{eq:Lnu}).
The transmission function may be expressed in terms of the junction Green's functions as \cite{Datta95,Bergfield09a}  
\begin{equation}
T_{\alpha\beta}(E)={\rm Tr}\left\{ \Gamma^\alpha(E) G^r(E) \Gamma^\beta(E) G^a(E)\right\},
\label{eq:transmission_prob}
\end{equation}
where $\Gamma^\alpha(E)$ is the tunneling-width matrix for lead $\alpha$
and $G^r(E)$ and $G^a$ are the retarded and advanced Green's functions of the junction, respectively.  
In the general many-body problem $G(E)$ must be approximated.  
In the context of the examples discussed here we consider an effective single-particle description such that
\begin{equation}
	G^r(E) = \left({\bf S}E-H - \Sigma^r_{T} \right)^{-1},
\end{equation}
where $H$ is the Hamiltonian of the nanostructure, ${\bf S}$ is an overlap matrix which reduces to the identity matrix in an orthonormal basis, 
and $\Sigma_{T}$ is the tunneling self-energy.

The tunneling-width matrix for contact $\alpha$ (source, drain, or probe) may be expressed as 
\begin{equation}
\left[\Gamma^\alpha(E)\right]_{nm} = 2\pi \sum_{k\in\alpha} V_{nk} V_{mk}^\ast\, \delta(E-\epsilon_k),
	\end{equation}
where $n$ and $m$ label $\pi$-orbitals within the graphene flake,
and $V_{nk}$ is the coupling matrix element 
between orbital $n$ of the graphene and a single-particle energy eigenstate of energy $\epsilon_k$ in electrode $\alpha$.
The thermal probe is treated as an ideal broad-band probe with perfect spatial resolution
\begin{equation}
\Gamma^p = \gamma_p \delta({\bf x}-{\bf x}_p),
\end{equation}
while the coupling to the hot and cold electrodes is taken to be diagonal in the graphene atomic basis
with a per-bond broad-band coupling strength of 3eV.
In the broad-band limit, the tunneling self-energy is a constant matrix given by
\begin{equation}
\Sigma^r_{T} = -\frac{i}{2} \sum_\alpha \Gamma^\alpha.
\end{equation}

In the low-energy regime (i.e., near the Dirac point), a simple tight-binding Hamiltonian has been shown to accurately describe the
$\pi$-band dispersion of graphene \cite{Reich02}.  
The Hamiltonian of the graphene flake is taken as
\begin{equation}
	H_{\rm graphene} = \sum_{\langle ij \rangle} t_{ij} d_i^\dagger d_j + {\rm H.c},
\end{equation}
where $t=-2.7\mbox{eV}$ is the nearest-neighbor hopping matrix element between 2p$_z$ carbon orbitals of the graphene flake with 
lattice constant of 2.5\AA, and $d_i^\dagger$ creates an electron on the i$^{th}$ 2p$_z$ orbital.  

The heat current density plotted in Fig.\ \ref{fig:heat_flux} is given within NEGF theory by
\begin{equation}
J_q = \frac{\hbar}{2m} \int \frac{dE}{2\pi} (E-\mu_0) \left(\nabla - \nabla^\prime\right) G^<({\bf x},{\bf x}^\prime; E),
\end{equation}
where $G^<$ is the Keldysh lesser Green's function (see, e.g., Ref.\ \onlinecite{stafford2016local} for its definition).


%

\end{document}